\documentclass[12pt]{article}

\usepackage[margin=1in]{geometry}
\usepackage{amsmath,amssymb,amsthm,amsfonts}
\usepackage{graphicx}
\usepackage{url}
\usepackage{booktabs}
\usepackage{multirow}
\usepackage{algorithm}
\usepackage{algpseudocode}
\usepackage[round,authoryear]{natbib}
\usepackage{setspace}
\usepackage{amsthm} 
\usepackage{authblk}
\usepackage{bm}
\usepackage[hidelinks]{hyperref}
\usepackage{tikz}
\usetikzlibrary{positioning,fit,calc,decorations.pathreplacing}
\usepackage{longtable}
\usepackage{float}
\usepackage{caption}
\usepackage{subcaption}

\bibpunct{(}{)}{;}{a}{,}{,}

\theoremstyle{plain}
\newtheorem{thm}{Theorem}

\theoremstyle{definition}
\newtheorem{asm}{Assumption}

\newtheorem{rem}{Remark}
\newtheorem{exm}{Example}

\allowdisplaybreaks

\title{\bf Evaluation of Combination Therapy amid Patient-Level Heterogeneity}

\author[1]{Lingjie Wei\thanks{Email: \texttt{lweiam@connect.ust.hk}}}
\author[1]{Xinzhou Guo\thanks{Corresponding author. Email: \texttt{xinzhoug@ust.hk}}}
\author[2]{Shuoxun Xu\thanks{Corresponding author. Email: \texttt{shuoxunxu\_ucb@berkeley.edu}}}

\affil[1]{Department of Mathematics\\
The Hong Kong University of Science and Technology\\
Hong Kong SAR, P.R.C.}

\affil[2]{Department of Biostatistics and Epidemiology\\
University of California\\
Berkeley, U.S.A.}

\date{\today}

\begin{document}

\maketitle

\bigskip
\begin{abstract}
Combination therapy, a treatment approach that involves two or more monotherapies, is widely considered to enhance therapeutic efficacy across different medical conditions. It was often believed that combination therapy is beneficial because of pharmacological interactions between its component monotherapies. However, through laboratory experiments, pharmacologists have recently noted that the benefits of some combination therapies might be largely driven by varying patient-level responses to their component monotherapies. Without accounting for such patient-level heterogeneity, classical statistical inference frameworks for combination therapy might be inadequate and overly optimistic. In this paper, we introduce a novel and model-free statistical inference framework to complement the classical one and evaluate combination therapy after adjusting for patient-level heterogeneity in responses to monotherapies. We address the non-identifiability and nonlinearity issue inherent in adjustment of patient-level heterogeneity and establish conditions for the (partial) identifiability of the cross-world target parameter. We develop an outcome-based optimal matching scheme to achieve asymptotic normality and construct $\sqrt{N}$-rate confidence intervals for the target parameter, thereby enabling reliable, efficient and transparent evaluation of combination therapy amid patient-level heterogeneity. The benefits of the proposed framework are demonstrated through a reanalysis of the ACTG 175 trial.
\end{abstract}

\noindent\textbf{Keywords:}  cross-world parameter; model-free; nonlinearity; outcome-based optimal matching; partial identification.
\bigskip

\section{Introduction}
\label{sec:intro}

In modern drug development, the study of combination therapy plays a central role. For many health conditions, a single therapy may have limited efficacy, while combination therapies exhibit potential benefits. For example, in treating HIV-infected patients, combination therapy with zidovudine plus didanosine was shown to be more effective than zidovudine alone \citep{hammer1996trial}. Over the past decades, many combination therapies were developed, and have been widely used in treating different types of conditions, such as cancer \citep{wei2024promise}, infectious diseases \citep{manosuthi2016integrated} and autoimmune disorders \citep{perini2008cyclophosphamide}. Recent data show that combination drugs account for 33.9\% of FDA cancer drug approvals and are more frequently approved as first-line treatments than monotherapies \citep{chen2025trends}. Therefore, understanding and validating the benefit of combination therapies has attracted sustained interest from both the pharmacological, statistical and regulatory communities.

Traditionally, the benefit of combination therapies has often been interpreted in terms of pharmacological interactions between its components, including both additive and synergistic effects \citep{palmer2017combination, plana2022independent}.\footnote{Terminology concerning drug interactions in combination therapy varies across the pharmacological literature. In line with the contrast discussed by \citet{palmer2017combination}, we use the term ``interaction effect'' in a broad sense that encompasses both additive and synergistic effects; see the Appendix for further discussion.} However, accumulating pharmacological evidence from both laboratory experiments and clinical data has revealed that the apparent efficacy of many combinations may be largely driven by patient-level heterogeneity of monotherapies. Here, patient-level heterogeneity refers to scenarios where a patient might respond differently to different monotherapies, and respond to the combination therapy based on the best monotherapy efficacy achieved for that patient, rather than through additional interaction from combining the monotherapies. This mechanism is also known as ``independent action'' or the ``highest single agent'' principle in the pharmacological literature \citep{plana2022independent, hwangbo2023additivity}. Classical evaluation of combination often ignores such patient-level heterogeneity and simply compares the effect size of the combination therapy with those of single agents, which may be inadequate and overly optimistic amid patient-level heterogeneity. 

A notable example illustrating the potential issues of classical evaluation of combination therapy amid patient-level heterogeneity is the study of melanoma. It was shown that the efficacy of ipilimumab plus nivolumab is greater than the efficacy of ipilimumab or nivolumab alone in treating melanoma, which, under the classical evaluation framework, may be taken as evidence that ipilimumab plus nivolumab has interaction effect and is beneficial. However, \citet{palmer2017combination} argues that this apparent benefit can be largely explained by patient-level heterogeneity of ipilimumab and nivolumab and the interaction effect might not truly exist. If a combination's benefit is driven by patient-level heterogeneity, classical evaluation may lead to inadequate and overly optimistic interpretations about the mechanism and the benefit of the combination therapy: (i) it may be inadequate and wrongly attribute the benefit of the combination therapy to drug--drug interaction, and (ii) it may be overly optimistic in assessing the added value of the combination as comparable outcomes may in principle be achieved by assigning the appropriate monotherapy to each patient and combination may induce additional treatment-related risks. In addition to melanoma, concerns that classical evaluation may not be able to separate interaction from patient-level heterogeneity also arise in other medical conditions, such as autoimmune diseases and HIV. These underscores the need for a new evaluation framework for combination therapies that adjusts for the best individual-level response to monotherapies and complements the classical one.

In this paper, we propose a novel evaluation framework for assessing  combination therapies amid patient-level heterogeneity. Specifically, under the potential outcome framework, we propose a new quantity to measure the treatment effect of the combination therapy after adjusting for individual-level maximum responses to monotherapies. The new quantity naturally accounts for patient-level heterogeneity, and thus provides an appropriate evaluation of combination therapy and prevents inadequate and overly optimistic findings in combination drug development amid patient-level heterogeneity. However, as the adjustment needs to be made at patient level, inherent issues of non-identifiability and non-linearity arise, creating fundamental challenges for inference on the proposed quantity of interest. In this paper, we establish the (partial) identifiability conditions and propose an outcome-based optimal matching algorithm to impute missing potential outcomes and approximate the maximal patient-level response to the monotherapies in a transparent way. We establish the asymptotics of the proposed matching estimator via a novel piecewise inverse-weighting argument in the presence of non-linearity of target functional, and show that the proposed method leads to valid and $\sqrt{N}$ inference of the quantity of interest under a model-free setup. We apply the proposed method to the HIV study, where combination therapy is often considered. Based on the ACTG 175 trial, the proposed method finds no evidence that the benefit of zidovudine plus didanosine exceeds the patient-level heterogeneity-driven effect, while a classical evaluation fails to distinguish such heterogeneity benchmark from interaction effect.

The evaluation of combination therapy under patient-level heterogeneity has gained increasing attention in both the statistical and pharmacological communities. Based on laboratory experiments, \citet{demidenko2019statistical} revisits specific combination therapies in in-vitro and in-vivo experiments. However, such lab-based methods typically adjust heterogeneity at the cell level, where cell copying is possible, and the results may not be easily extrapolated to human clinical settings \citep{sackett1989rules}. To account for patient-level heterogeneity in clinical settings, several methods have been proposed for evaluating combination therapy, such as the resampling-based approaches by \citet{palmer2017combination} and the model-based frameworks by \citet{chen2020independent}. However, these methods are either ad-hoc or model-based, and more importantly, do not appropriately define and study the treatment effect of the combination therapy after accounting for patient-level heterogeneity under a rigorous potential-outcome-based framework. We refer to \cite{schmidt2023rationales} and \cite{plana2022independent} for a more comprehensive review of the existing literature. 

Our work is also related to, yet substantially different from, several research protocols in precision medicine and causal inference. From an application perspective, our framework is related to the studies of optimal treatment regimes \citep{luedtke2016statistical, xu2025optimal} as both concern treatment effect heterogeneity. However, rather than identifying the best treatment regimes for different patient profiles, our work aims to adjust for patient-level heterogeneity in combination therapy evaluation over the overall population from a pharmacological and mechanistic perspective, leading to a distinct causal quantity of interest. Similar differences also arise between our framework and the studies of individual treatment effects \citep{lei2021conformal, chernozhukov2023toward} and cross-world causal parameters \citep{wu2024quantifying, wu2025promises, bodik2025cross}, which also involve joint features of potential outcomes but mainly focus on individual-level prediction or cross-world identification of categorical outcomes given multiple trials, rather than inference on combination therapy efficacy in a clinical trial. From a methodological perspective, our framework is related to matching literature \citep{abadie2006large, abadie2016matching,yang2023multiply}. However, existing matching methods are often based on covariates and propensity scores and aim for a linear quantity, while we propose an outcome-based matching and the quantity of interest is non-linear.
 
In summary, our contributions are threefold: framework, methodology, and theory. First, we propose a new evaluation framework for combination therapy, accounting for patient-level heterogeneity under the potential outcome framework, and establish (partial) identifiability conditions for the target quantity. Second, we develop an outcome-based optimal matching procedure to impute potential outcome, enabling transparent and valid $\sqrt{N}$ inference for the target quantity, which is nonlinear. Third, we employ a novel piecewise inverse-weighting argument and provide a novel route for deriving asymptotic properties of matching estimators for nonlinear estimands.

The rest of the paper is organized as follows. In Section~\ref{section:setting}, we introduce the problem setting, discuss the limitation of the classical evaluation framework of combination therapy amid patient-level heterogeneity, and formulate the new evaluation framework adjusting for patient-level heterogeneity under the potential outcome framework. In Section~\ref{section:method}, we establish the identification conditions and introduce the matching estimator, showing that it achieves asymptotic normality and enables a $\sqrt{N}$ sharp inference of the proposed quantity of interest under the identifiablity conditions. In Section~\ref{section:robustness}, we study the robustness of the proposed evaluation framework, establishing partial identification conditions and proposing a sensitivity analysis framework. Section~\ref{section:sim} demonstrates the validity and robustness of our methods across different simulation settings. Finally, Section~\ref{section:application} re‑analyzes the ACTG 175 trial and evaluates the treatment effect of zidovudine plus didanosine amid patient‑level heterogeneity.

\section{Problem Setting and Evaluation Framework}\label{section:setting}
In this section, we first present the problem setting for evaluating combination therapy. Subsequently, we revisit traditional evaluation approaches and illustrate their limitations amid patient-level heterogeneity through a toy example. Finally, we develop a novel framework that adjusts the patient-level heterogeneity in combination therapy evaluation and motivates the matching algorithm in the next section.

\subsection{Problem Setting and Classical Evaluation}

Consider a clinical trial of $N$ patients with the observed data $\mathcal{I} = \{(Y_i, \mathbf{X}_i, T_i)\}_{i=1}^N$, an independent and identically distributed copy of $(Y, \mathbf{X}, T)$. Here, $Y$ is the observed outcome, $T$ is the treatment assignment, and $\mathbf{X}$ is a $k$-dimensional baseline covariate, i.e., $\mathbf{X} \in \mathbb{X} \subseteq \mathbb{R}^k$. Without loss of generality, we consider the combination therapy consists of two drug agents, and let $T$ denote treatment, which takes values in $\{A, B, A+B\}$, corresponding to monotherapy $A$, monotherapy $B$, and combination therapy $A+B$, respectively. Under the potential outcome framework, the observed outcome $Y$ is defined as $$Y = \mathbb{I}\{T = A\} Y(A) + \mathbb{I}\{T = B\} Y(B) + \mathbb{I}\{T = A+B\} Y(A+B),$$ where $(Y(A), Y(B), Y(A+B))$ denote the potential outcomes under treatments $A$, $B$, and $A+B$, and $\mathbb{I}\{\cdot\}$ is the indicator function. Throughout the paper, we assume that the treatment assignment is completely randomized, i.e. $(Y(A), Y(B), Y(A+B), \mathbf{X}) \perp T$, in the clinical trial.

Classical evaluation of combination therapy typically compares the average treatment effect of combination therapy with those of monotherapies it consists of directly (see, e.g. \cite{schmidt2020assessment,schmidt2023rationales,duarte2022evaluation}) to study the mechanism and the benefit of the combination therapy. Without loss of generality, such classical evaluation can be formulated as the following hypothesis test,
\begin{equation}\label{basic}
H_0: E[Y(A+B)] \leq \max\{E[Y(A)], E[Y(B)]\} 
\quad \text{vs.} \quad 
H_1: E[Y(A+B)] > \max\{E[Y(A)], E[Y(B)]\},     
\end{equation}
where the alternative hypothesis implies that the treatment effect of the combination therapy is superior to the treatment effect of any monotherapy it consists of at the population level. When the null hypothesis in Eq. \eqref{basic} was rejected, some researchers used to anticipate that (1) a pharmacological interaction between treatments A and B exists and (2) the combination therapy of $A+B$ is beneficial with significant added values, which nevertheless, based on the recent development in pharmacology, might not be exactly true due to the ignoring of patient-level heterogeneity in classical evaluation.

As recently noted by pharmacologists \citep{palmer2017combination}, evaluation of the combination therapy by simply testing the hypothesis in Eq. \eqref{basic} might be inadequate and overly optimistic. Specifically, rejecting the null hypothesis in Eq. \eqref{basic} does not necessarily imply a pharmacological interaction between the two agents of the combination therapy because of the potential existence of the patient-level heterogeneity. In pharmacology, patient-level heterogeneity refers to the scenario that a patient might respond differently to different monotherapies and respond to combination therapy by the best monotherapy efficacy it constitutes for that patient, rather than by additional interaction from combining the monotherapies. Pharmacologists argue that sometimes, the so-called mechanism and beneficiary of the combination therapy implied by the classical evaluation in Eq. \eqref{basic} might be indeed largely driven by the patient-level heterogeneity of the monotherapy, and  thus might not be well-grounded enough.   

To translate and fix the above pharmacological idea in a statistical way, we let $E[\max\{Y(A), Y(B)\}]$ denote such patient-level heterogeneity efficacy that each patient achieves the best individual response to treatments A and B. By Jensen's inequality, unless patient-level heterogeneity does not exist; i.e. $Y(A)\ge Y(B)$ or $Y(A)<Y(B)$, a.s., $E[\max\{Y(A), Y(B)\}]> \max\{E[Y(A)], E[Y(B)]\}$. Therefore, the alternative in Eq. \eqref{basic} can hold even when 
\begin{equation}\label{inequal}
    E[\max\{Y(A), Y(B)\}]\ge E[Y(A+B)] >\max\{E[Y(A)], E[Y(B)]\}.
\end{equation}
When this happens, it suggests similar treatment effects may be achieved in principle if precision medicine is applicable; i.e. assigning correct monotherapy to the right patient, rather than combination therapy. Therefore, amid patient-level heterogeneity, even when the null hypothesis in Eq.~\eqref{basic} is rejected, it may not necessarily imply that (1) a pharmacological interaction between the single agents exists or that (2) the combination therapy provides additional benefit beyond the patient-level best monotherapy, particularly given the potentially increased risk of using more than one agent simultaneously. These differ from the anticipation only based on the classical evaluation in Eq.~\eqref{basic}.

Note that patient-level heterogeneity is different from the optimal treatment regime (OTR). Specifically, the mean value of the OTR, $E[\max\{E[Y(A)|\mathbf{X}], E[Y(B)|\mathbf{X}]\}]$, relies on $\mathbf{X}$ for its formulation, while the patient-level heterogeneity benchmark considered in this paper, $E[\max\{Y(A), Y(B)\}]$, is defined without $\mathbf{X}$. This is due to the different natures of the scientific problems they aim to address \citep{palmer2017combination} that patient-level heterogeneity is studied in pharmacological study and pharmaceutical development to appropriately interpret the mechanism and quantify the benefit of the combination therapy in the overall population, rather than to find the optimal treatment regime for different patient's profile. 

In summary, there are two potential limitations of the classical evaluation framework in Eq. \eqref{basic}. First, it fails to distinguish a pharmacological interaction effect from the benefit driven purely by patient-level heterogeneity, which is represented by $E[\max\{Y(A), Y(B)\}]$. Second, it overlooks the possibility that the efficacy of the combination therapy is similar to $E[\max\{Y(A), Y(B)\}]$ and thus may be achieved in principle by assigning monotherapies precisely. When this happens, both drug developers and regulatory agencies should be more cautious, as the anticipated mechanistic rationale and added value of pharmacological interaction in combination therapy may become questionable, especially given the possible additional risks of the combination therapy \citep{nesto2003thiazolidinedione}. Thus, an appropriate adjustment for patient-level heterogeneity is needed in the evaluation of combination therapy to better understand its underlying mechanism and inform clinical or regulatory decision-making of combination therapy. To conclude this section, we provide a toy illustration in Example~\ref{Example1}.

\begin{exm}\label{Example1}
Consider a toy example where male patients have higher response to Drug A ($Y(A)=10$) than to Drug B ($Y(B)=5$), and female patients have higher response to Drug B ($Y(B)=10$) than to Drug A ($Y(A)=5$). Suppose the potential outcome under combination therapy equals the best monotherapy response, i.e., \(Y(A+B)=\max\{Y(A),Y(B)\}\); thus there is no true interaction between the two drugs. Under equal gender proportions, this patient-level heterogeneity yields $E[\max\{Y(A),Y(B)\}]=10$, which strictly exceeds the conventional threshold $\max\{E[Y(A)],E[Y(B)]\}=7.5$ as shown in Figure~\ref{fig:heterogeneity}. Consequently, in this setting, the classical evaluation in Eq.~\eqref{basic} may still misinterpret this gap as an additional benefit and suggest a pharmacological interaction, whereas the benefit of $A+B$ is purely driven by patient-level heterogeneity and could, in principle, be attained by an oracle assignment of drug A to males and drug B to females.
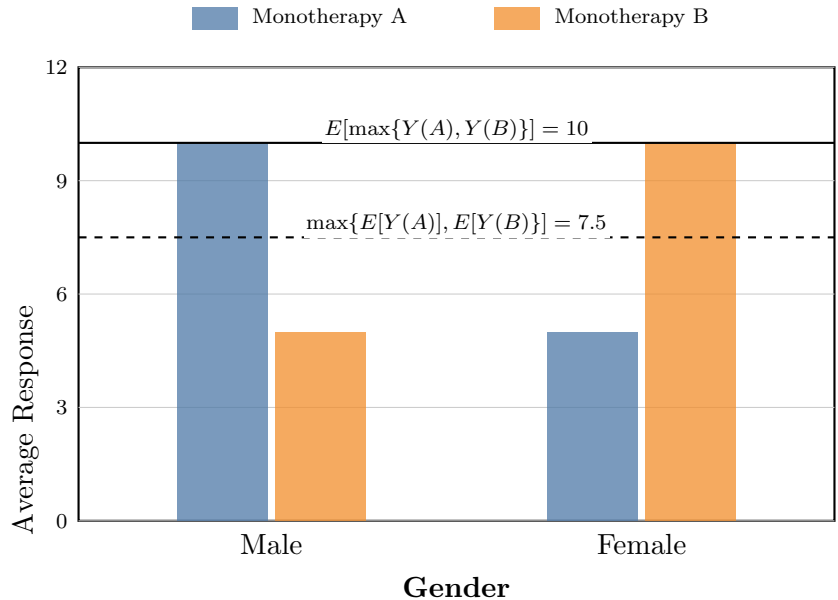
\begin{figure}[H]
\centering
\begin{tikzpicture}[>=stealth, yscale=0.5]

\definecolor{colorA}{RGB}{78,121,167}   
\definecolor{colorB}{RGB}{242,142,44}   

\draw[thick] (0,0) rectangle (10,12);
\foreach \y/\val in {0/0, 3/3, 6/6, 9/9, 12/12}{
    \draw[thin, gray!40] (0,\y) -- (10,\y);
    \node[left, font=\scriptsize] at (0,\y) {\val};
}
\node[left, font=\small, rotate=90] at (-0.7, 6) {Average Response};

\fill[colorA, opacity=0.75] (1.3, 0) rectangle (2.5, 10.0);
\fill[colorB, opacity=0.75] (2.6, 0) rectangle (3.8, 5.0);

\fill[colorA, opacity=0.75] (6.2, 0) rectangle (7.4, 5.0);
\fill[colorB, opacity=0.75] (7.5, 0) rectangle (8.7, 10.0);

\node[font=\small] at (2.55, -0.6) {Male};
\node[font=\small] at (7.45, -0.6) {Female};
\node[font=\small, below] at (5.0, -1.2) {\textbf{Gender}};

\draw[thick, black] (0,10.0) -- (10,10.0);
\node[font=\scriptsize, fill=white, inner sep=1pt] at (5.0, 10.35) {$E[\max\{Y(A),Y(B)\}] = 10$};

\draw[thick, black, dashed] (0,7.5) -- (10,7.5);
\node[font=\scriptsize, fill=white, inner sep=1pt] at (5.0, 7.85) {$\max\{E[Y(A)],E[Y(B)\}] = 7.5$};

\fill[colorA, opacity=0.75] (1.5, 13.0) rectangle (2.1, 13.6);
\node[right, font=\scriptsize] at (2.15, 13.3) {Monotherapy A};
\fill[colorB, opacity=0.75] (5.5, 13.0) rectangle (6.1, 13.6);
\node[right, font=\scriptsize] at (6.15, 13.3) {Monotherapy B};
\end{tikzpicture}
\caption{Illustration of patient-level heterogeneity. The solid horizontal line indicates the joint potential outcome benchmark $E[\max\{Y(A),Y(B)\}]=10$, while the dashed horizontal line indicates the classical monotherapy comparison threshold $\max\{E[Y(A)],E[Y(B)]\}=7.5$.}
\label{fig:heterogeneity}
\end{figure}
\end{exm}

\subsection{Proposed Evaluation Framework}

In this paper, we propose a new statistical inference framework to evaluate combination therapy after adjusting for patient-level heterogeneity of monotherapies. In particular, we consider the following hypothesis test
\begin{equation}\label{basic2}
    H_0:\tau\leq0\quad \text{vs.} \quad H_1:\tau>0
\end{equation}
where $\tau=E[Y(A+B)]-E[\max\{Y(A), Y(B)\}]$ is the average treatment effect of combination therapy after subtracting the mean of the best individual-level response to monotherapies A and B. Therefore, failing to reject the null hypothesis in Eq. \eqref{basic2}; i.e. $E[Y(A+B)]\leq E[\max\{Y(A), Y(B)\}]$, implies that the benefit of combination therapy might not be driven by a pharmacological interaction \citep{palmer2017combination}, but by the independent action driven by individual-level heterogeneity. Such a result suggests that drug developers and regulatory agencies should more carefully assess the added value and further adoption of the combination therapy $A+B$, especially when the inclusion of more agents may increase the toxicity burden \citep{us2013codevelopment}.  On the contrary, rejecting the null hypothesis in Eq.~\eqref{basic2} supports that the benefit of combination therapy is attributed to a pharmacological interaction effect, and helps further justify the necessity of the development and adoption of combination therapy.

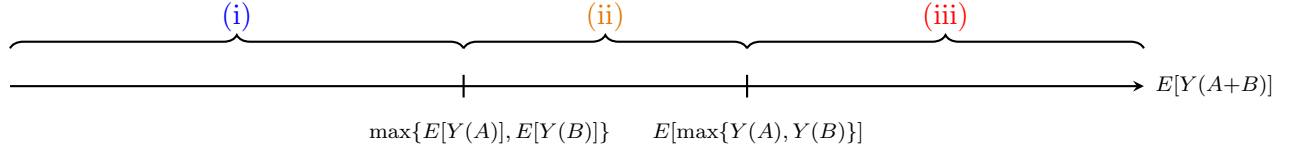
\begin{figure}[ht]
\centering
\tikzset{
    axis/.style={->, thick},
    tick/.style={thick},
    brace/.style={decorate,decoration={brace,amplitude=5pt}, thick},
    lbl/.style={midway, yshift=12pt, font=\small}
}
\begin{tikzpicture}[xscale=1.5, yscale=1.0, >=stealth]
\draw[axis] (-4,0) -- (6,0) 
    node[anchor=west, font=\scriptsize] {$E[Y(A{+}B)]$};
\draw[tick] (0,0.15) -- (0,-0.15)
    node[below=5pt, font=\scriptsize, align=center, text width=2.5cm] 
    {$\max\{E[Y(A)], E[Y(B)]\}$};
\draw[tick] (2.5,0.15) -- (2.5,-0.15)
    node[below=5pt, font=\scriptsize, align=center, text width=2.5cm] 
    {$E[\max\{Y(A), Y(B)\}]$};
\draw[brace] (-4,0.5) -- (0,0.5)
    node[lbl, text=blue]{\textcolor{blue}{(i)}};
\draw[brace] (0,0.5) -- (2.5,0.5)
    node[lbl, text=orange!90!black]{\textcolor{orange!90!black}{(ii)}};
\draw[brace] (2.5,0.5) -- (6,0.5)
    node[lbl, text=red]{\textcolor{red}{(iii)}};
\end{tikzpicture}
\caption{Categories of combination efficacy: \textcolor{blue}{(i)} not effective, 
\textcolor{orange!90!black}{(ii)} patient-level heterogeneity, 
\textcolor{red}{(iii)} true pharmacological interaction. Classical evaluation framework mistakenly groups regions (ii) and (iii) together.}
\label{fig:proposed_eval}
\end{figure}

Compared with Eq.~\eqref{basic}, Eq.~\eqref{basic2} helps identify interaction effect and quantify the benefit of combination therapy amid patient-level heterogeneity appropriately. As illustrated in Figure~\ref{fig:proposed_eval}, the recent development of pharmacologies suggests the treatment effect of the combination therapy indeed falls into three categories: (i) not effective, (ii) patient-level heterogeneity, and (iii) pharmacological interaction. The classical framework in Eq. \eqref{basic} indeed can only distinguish (ii) and (iii) as a whole from (i), and thus can not really identify true pharmacological interaction effect only in (iii), in Figure~\ref{fig:proposed_eval}. On the contrary, the proposed framework in Eq. \eqref{basic2} can distinguish (iii) from (i) and (ii), and thus identify a true pharmacological interaction beyond the patient-level heterogeneity benchmark, as illustrated in Figure~\ref{fig:proposed_eval}. It is in this sense that our proposed test in Eq. \eqref{basic2} complements, though not replace, classical superiority comparisons in Eq. \eqref{basic} by probing this mechanistic boundary and quantifying the true beneficiary of combination therapy.

Although the introduction of $\tau$ successfully accounts for patient-level heterogeneity in combination therapy evaluation, directly testing Eq.~\eqref{basic2} is challenging due to issues of non-identifiability and non-linearity. These challenges are inherent to the statistical question translated from pharmacology itself, as the adjustment for heterogeneity needs to be made at the patient level. First, unlike the average treatment effect (ATE) or the average treatment effect on the treated (ATT), which are identifiable based on marginal distributions (i.e., via the G-formula) under classical identification assumptions, $\max\{Y(A),Y(B)\}$ depends on the joint distribution of $(Y(A),Y(B))$ and is not explicitly identified even in a clinical trial. Unlike laboratory experiments where cell copying is possible, the pair $(Y(A),Y(B))$ is not jointly observed. Thus, new identification conditions need to be derived for $\tau$. Second, even setting aside identifiability, constructing transparent estimators and deriving their asymptotic properties for $\tau$ remain challenging because $E[\max\{Y(A),Y(B)\}]$ constitutes a non-linear functional. Unlike existing approaches relying on parametric assumptions, we impose no parametric structure on the joint distribution, making it infeasible to reduce the estimation of $\tau$ into a standard parameter estimation problem. Therefore, new estimation strategies and theory need to be developed. In the next section, we first address an identification issue that has been largely overlooked in the existing combination therapy literature and then introduce our proposed algorithm and its asymptotic properties.

\section{Methodology and Theory}\label{section:method}
In this section, we introduce the identification condition for the parameter of interest, propose an algorithm based on optimal matching on the estimated conditional outcome, and demonstrate the asymptotic normality and $\sqrt{N}$ inference validity of the proposed evaluation framework.

\subsection{Identification}\label{section:identification}

For notational convenience, let $\mu_{A+B} := E[Y(A+B)], \psi := E[\max\{Y(A), Y(B)\}],$ and thus $\tau := \mu_{A+B} - \psi.$ We consider the decomposition $Y_i(T_i) = f_{T_i}(\mathbf{X}_i) + \epsilon_i(T_i)$ with the noise term $E[\epsilon_i(T_i)|\mathbf{X}_i,T_i]=0$ and $f_{T_i}(\mathbf{X}_i)=E[Y_i(T_i)|\mathbf{X}_i]$. We assume the noise term satisfies certain homoscedasticity and regularity as stated in Assumption~\ref{assump:noise}, which holds in many practical scenarios, such as Gaussian noise with constant variance. Note that here we do not impose any functional form for $f_T(\cdot)$, and it is in this sense that we call the method proposed later model-free.

Inspired by the dependence modeling in \cite{chen2020independent}, we define a conditional dependence measure $Cov[\mathbb{I}_{Y(A)> t},\mathbb{I}_{Y(B)> t}|\mathbf{X}]$ and propose the following identification condition.

\begin{asm}[Conditional Bliss Independence]
\label{assump:bliss}
$Cov[\mathbb{I}_{Y(A)> t},\mathbb{I}_{Y(B)> t}|\mathbf{X}]=0$ for all $t\in\mathbb{R}\ \text{a.s.}.$
\end{asm}

Assumption~\ref{assump:bliss} can be viewed as a statistical adaptation of the Bliss independence, a widely used concept in pharmacology \citep{bliss1939toxicity, demidenko2019statistical}. Indeed, by conditioning on $\mathbf{X}$, it extends this classical idea to a conditional scenario, serving as an interpretable and realistic working baseline for causal identification of $\tau$.  Specifically, Assumption~\ref{assump:bliss} can be satisfied by the conditional cross-world independence between the noises of two potential outcomes, i.e., $\epsilon(A)\perp \epsilon(B)$ \citep{bodik2025cross}. Moreover, even if Assumption~\ref{assump:bliss} is violated, partial identification of the quantity of interest $\tau$ can still be established under broad practical scenarios. Detailed discussions on partial identification and sensitivity analysis regarding this baseline assumption can be found in Sections~\ref{section:partial} and \ref{section:sensitivity}.

Let $H$ denote the joint distribution of $(Y(A),Y(B),\mathbf{X})$ induced by the conditional product of the marginal measures $F_{Y(A)\mid\mathbf{X}}\times F_{Y(B)\mid\mathbf{X}}$ and the marginal distribution $F_{\mathbf{X}}$. We define 
\begin{equation}\label{psi_H}
 \psi_H := E_H[\max\{Y(A),Y(B)\}]
    = E_{F_\mathbf{X}}\!\left[
        E_{F_{Y(A)|\mathbf{X}}\times F_{Y(B)|\mathbf{X}}}\!\left(
            \max\{Y(A),Y(B)\} \mid \mathbf{X}
        \right)
    \right],    
\end{equation}
and $\tau_H:=\mu_{A+B}-\psi_H$, both of which are identifiable because they only depend on the functional of marginal distributions, i.e. $F_\mathbf{X},F_{Y(A)|\mathbf{X}}$ and $F_{Y(B)|\mathbf{X}}$. Under the conditional bliss independence in Assumption \ref{assump:bliss}, Theorem \ref{thm:identification} establishes identification of $\tau$ by expressing $\psi$ as $\psi_H$. We refer the partial identifiability of $\tau$ to Section~\ref{section:partial} when the conditional bliss independence is violated.

\begin{thm}[Identification]
\label{thm:identification}
Under Assumption~\ref{assump:bliss},  $\psi=\psi_H,\quad \tau=\tau_H=\mu_{A+B}-\psi_H$. Consequently, $\psi$ and $\tau$ are identified.
\end{thm}

Having established identifiability, however, our evaluation framework still faces two challenges. First, the identification formula in Eq.~\eqref{psi_H} depends on the full conditional laws $F_{Y(A)\mid\mathbf{X}}$ and $F_{Y(B)\mid\mathbf{X}}$ through the nonlinear $\max$ operator, so a naive plug-in method that directly plugs in empirical conditional distributions of the potential outcomes may fail to achieve the $\sqrt{N}$ rate. Second, how to approximate or impute the missing part of $\max\{Y_i(A),Y_i(B)\}$ in a transparent way is unclear due to the nonlinear $\max$ operator.  These challenges motivate the matching-based procedure developed below.

\subsection{Methodology}\label{sub:method}

In this subsection, we propose an optimal matching estimator based on the estimated conditional expectation of the outcome as summarized in Algorithm~\ref{alg1}. Here, without loss of generality, we assume $N_A\leq N_B$, while the algorithm can be naturally adapted when $N_B\leq N_A$. The key idea of Algorithm~\ref{alg1} is to match on the estimated conditional expectation of $Y(B)$, and impute the missing potential outcome $Y_i(B)$ for each subject $i$ in group $A$ using the observed outcome $Y_{\widetilde{i}}$ of their matched counterpart $\widetilde{i}$ from group $B$. This yields a directly computable $\max\{Y_i(A),Y_{\widetilde{i}}(B)\}$ that mimics $\max\{Y_i(A),Y_i(B)\}$ in a transparent way under the product measure $F_{Y(A)|\mathbf{X}_i}\times F_{Y(B)|\mathbf{X}_i}$. 

Algorithm~\ref{alg1} consists of three key steps.  First, Algorithm~\ref{alg1} starts with an outcome-based optimal matching, which matches subjects receiving treatment $A$ with subjects receiving treatment $B$ based on $\widehat{f}_{B}(\cdot)$, an estimator for the conditional outcome $f_{B}(\cdot)$. Here, the matching is based on the conditional outcome in order to mimic the independent draw from $F_{Y(B)|\mathbf{X}_i}$. Second, we impute $Y_i(B)$ with the matched subject's outcome $Y_{\widetilde{i}}$. This allows us to construct a transparent mimic $\max\{Y_{i}, Y_{\widetilde{i}}\}$ for $\max\{Y_{i}(A),Y_{i}(B)\}$ and thus the estimator for $\tau$.  In the end, Algorithm~\ref{alg1} calculates a Wald-type confidence interval and $p$-value based on a two-sample t-test with the plug-in standard error.

\begin{algorithm}[H]
\caption{Estimation and inference for treatment effect of combination therapy after adjusting for patient-level heterogeneity via outcome-based optimal matching}
\label{alg1}
\small
\textbf{Input:} Clinical trial data $\mathcal{I}=\{Y_i,\mathbf{X}_i,T_i\}_{i=1}^{N}$ and an externally estimated outcome model $\widehat{f}_B(\mathbf{X})$ for $E[Y(B)\mid\mathbf{X}]$.

\smallskip
\textbf{Target parameter:} $\tau = E[Y(A+B)] - E[\max\{Y(A), Y(B)\}]$.\\

\begin{algorithmic}[1]
\State \textbf{Optimal 1-to-1 matching:} Mimic independent draws from $F_{Y(B)\mid\mathbf{X}}$ by solving:
\[
\{\widetilde{i}:T_{i}=A\} = \arg\min_{j\in\mathcal{I}_B}\;
\sum_{i\in\mathcal{I}_A}\bigl|\widehat{f}_B(\mathbf{X}_{i})-\widehat{f}_B(\mathbf{X}_{j})\bigr|
\]

\State \textbf{Matching estimation:} Impute potential outcomes and compute the matching estimator $\widehat{\tau}$:
\[
\widehat{\tau} = \frac{1}{N_{A+B}} \sum_{T_i = A+B} Y_i \;-\; \frac{1}{N_A} \sum_{T_{i} = A} \max\{ Y_{i},\, Y_{\widetilde{i}} \}
\]

\State \textbf{Variance estimation, confidence interval, and hypothesis test:} Calculate the plug-in standard error $\widehat{\sigma} = \sqrt{\widehat{\sigma}^2}$ via:
\[
\widehat{\sigma}^2 = \frac{\widehat{\sigma}^2_{A+B}}{N_{A+B}} + \frac{\widehat{\sigma}^2_{m}}{N_{A}}
\]
where $\widehat{\sigma}^2_{A+B}$ and $\widehat{\sigma}^2_{m}$ are the sample variances of $\{Y_i:T_i=A+B\}$ and $\{\max\{Y_{i},Y_{\widetilde{i}}\}:T_{i}=A\}$, respectively. Construct the one-sided confidence interval and compute the $p$-value:
\[
CI_N = \bigl[ \widehat{\tau} - z_{1-\alpha}\,\widehat{\sigma},\; +\infty \bigr), \qquad p_N = 1 - \Phi\left(\frac{\widehat{\tau}}{\widehat{\sigma}}\right)
\]
where $z_{1-\alpha}$ is the $(1-\alpha)$-quantile of the standard normal distribution and $\Phi(\cdot)$ is the standard normal CDF.
\end{algorithmic}
\end{algorithm}

Although there is a rich literature on matching methods \citep{abadie2006large,abadie2016matching,abadie2022robust,hansen2008prognostic} and their asymptotic theory \citep{abadie2012martingale,antonelli2018doubly,yang2023multiply}, our matching-based imputation approach differs from existing matching methods both methodologically and theoretically.  

Methodologically, our matching scheme is outcome-based. Matching has been widely used in the design stage of observational studies as a tool to reduce imbalance, often being covariate-based and propensity-score-based \citep{rosenbaum2010design,stuart2010matching,wu2024matching}. In our clinical setting, however, covariate balance is achieved by design, so matching is not used for confounding adjustment. Instead, it is employed to impute the unobserved outcome-related quantity in a transparent way, which naturally calls for a matching score that summarizes outcome-relevant information as we did in Algorithm \ref{alg1}. Specifically, our proposed scheme performs matching on the one-dimensional estimated outcome score $\widehat{f}_B(\mathbf{X})$, which serves as a robust dimension-reduction tool and helps mitigate the issue of curse of dimensionality suffered by direct matching on the multivariate covariates $\mathbf{X}$ \citep{abadie2006large, ferman2021matching}. 

Theoretically, the proposed matching estimator targets $\psi = E[\max\{Y(A), Y(B)\}]$, which is a non‑linear estimand. A key challenge to establish $\sqrt{N}$‑asymptotics of the matching estimator is that matching inherently introduces dependence between matched pairs, invalidating the classical central limit theorems \citep{abadie2006large,abadie2012martingale,abadie2022robust}. To address this dependence issue, classical asymptotic analyses of matching estimators are often built upon linearization techniques, typically relying on $K(i)$, the number of times unit $i$ is matched. However, such approaches are not applicable here due to the non‑linearity of $\psi$ \citep{abadie2006large,lin2023estimation}. Another theoretical pathway is Le Cam's framework, which has been applied to parametric or semi‑parametric models involving prognostic or predictive scores \citep{rubin1986statistical,antonelli2018doubly,yang2023multiply}. Yet, these results rely on parametric modeling assumptions and thus are not suitable for the model‑free setup considered here. The next subsection develops a novel theoretical framework that addresses these challenges and justifies Algorithm \ref{alg1}. To conclude this subsection, we discuss the detailed implementation of Algorithm \ref{alg1} in Remark \ref{rm}.

\begin{rem}\label{rm}
Two implementation details of Algorithm~\ref{alg1} warrant further discussion. First, the estimated outcome model $\widehat{f}_B$ is supplemented externally, which aims to mitigate dependency issues induced by potentially complex working models of $\widehat{f}_B$. External estimation of the matching score is often assumed in the theoretical analysis of matching estimators and is feasible in many clinical studies. Specifically, the evaluation of combination therapy often involves an approved agent \citep{us2013codevelopment}, whose earlier-phase historical data may be used to provide the external estimated outcome model $\widehat{f}_B$ \citep{czechowicz2025old,freidlin2023augmenting}. Moreover, as $\widehat{f}_B$ only needs to satisfy a mild nonparametric rate in Assumption \ref{assump:unif-ext}, we might split the dataset $\mathcal{I}$ to get the external estimated outcome model $\widehat{f}_B$ without a real external dataset. In practice, one may directly use the whole dataset $\mathcal{I}$ to provide the estimated outcome model $\widehat{f}_B$, and our simulation results in Section~B of Appendix suggest that such an internal method remains empirically robust. Second, the matching scheme is based on optimal matching, which aims to minimize matching bias in the one-dimensional estimated outcome score required in Assumption \ref{assump:match-bias}. In practice, one may simply use the nearest-neighbor matching algorithm, and our simulation results in the Appendix suggest that standard matching remains empirically robust.
\end{rem}

\subsection{Theoretical Justifications}
In this subsection, we develop a new theoretical route to establish $\sqrt{N}$‑asymptotics of the proposed matching estimator in Algorithm \ref{alg1} for the non-linear quantity of interest $\tau$ under the model-free setup. The core idea is to construct a pseudo‑oracle quantity that is independent across matched pairs and then bound the approximation error between the pseudo-oracle quantity and the matched quantity used by the proposed estimator under mild conditions on estimation error and matching bias. This new strategy is substantially different from the usual use of $K(i)$ and Le Cam's theory in existing theoretical analysis of matching estimator, which are nevertheless not trivially applicable here as the estimand $\tau$ is non-linear. 

To justify Algorithm \ref{alg1}, we first impose Assumption \ref{assump:noise}, which requires the noise satisfy a mild independent and continuous structure. The independent structure, i.e., the noise distribution does not depend on the covariate \(\mathbf{X}\), has been widely adopted in the theoretical analyses of matching estimator; see, for example, \cite{han2023noisy,deng2021confidence}. The continuity condition on the noise distribution can be satisfied by many common noise distributions, such as the Gaussian and Laplace distributions, in practice.

\begin{asm}[Regularity on noises]
\label{assump:noise} 
Consider the model $Y_i(T_i) = f_{T_i}(\mathbf{X}_i) + \epsilon_i(T_i)$, where $\epsilon_i(T)$ are i.i.d. copies of $\epsilon(T)$ such that $\mathbf{X}\perp \epsilon(T)$ and $\epsilon(T)$ is absolutely continuous w.r.t. the Lebesgue measure on $\mathbb{R}$ and has a bounded probability density.
\end{asm}

The next two assumptions are imposed to bound two error sources, namely estimation error and matching bias, which are requisite for $\sqrt{N}$-consistency and the subsequent asymptotic results. Specifically, Assumption~\ref{assump:unif-ext} ensures a nonparametric estimation accuracy of the outcome model, while Assumption~\ref{assump:match-bias} guarantees that the matching discrepancy vanishes at a reasonable pace.

\begin{asm}[Uniform convergence of outcome model estimation]
\label{assump:unif-ext}
  $\|\widehat{f}_B-f_B\|_\infty=o_P(N_A^{-1/4})$, where $\|g\|_\infty=\sup_x|g(x)|$ is the supremum norm of function $g$.
\end{asm}

\begin{asm}[Matching bias]
\label{assump:match-bias}
    $E[\widehat{f}_B(\mathbf{X}_{i})-\widehat{f}_B(\mathbf{X}_{\widetilde{i}})]^4=o_P(N_A^{-1})$, where $i$ can be any index in $\{i:T_i=A\}$ due to the exchangeability of optimal matching.
\end{asm}

Assumption~\ref{assump:unif-ext} concerns the uniform estimation accuracy of the outcome model $\widehat{f}_B$ at the rate of $o_P(N_A^{-1/4})$. This is a mild nonparametric rate that can be satisfied by many nonparametric estimators, e.g. Newey's series estimator \citep{abadie2011bias,lin2023estimation}, while classical score-based matching theories typically rely on correctly specified parametric model in order to achieve $\sqrt{N}$ inference \citep{abadie2016matching,antonelli2018doubly,yang2023multiply}. The nonparametric rate requirement in Assumption~\ref{assump:unif-ext} not only permits a more flexible estimation of $\widehat f_B$ but even allows us to estimate $\widehat{f}_B$ by sample split of $\mathcal{I}$ as discussed in Remark \ref{rm}. 

Assumption~\ref{assump:match-bias} controls the matching discrepancy between each treated unit and its matched control, measured through the estimated outcome model. This type of assumption is typically required in asymptotic analyses of matching estimators. Moreover, compared with the classical matching bias assumption, such as Assumption~3 in \cite{abadie2022robust}, Assumption~\ref{assump:match-bias} is weaker as it only requires $o(N_A^{-1/4})$ in terms of the first-order difference, while $o(N_A^{-1/2})$ is imposed in \cite{abadie2022robust}.  Under the optimal matching scheme, the practical relevance and the sufficient conditions for Assumption~\ref{assump:match-bias} can be found and adapted from Proposition 1 in \cite{abadie2012martingale}.

Together, Assumptions~\ref{assump:bliss}, \ref{assump:noise}, \ref{assump:unif-ext}, and~\ref{assump:match-bias} characterize the theoretical framework under which our estimator exhibits valid and $\sqrt{N}$ asymptotic behavior as summarized in Theorems \ref{thm:normality} and \ref{thm:inference}. Specifically, Theorem~\ref{thm:normality} demonstrates the $\sqrt{N}$-asymptotic normality of the proposed matching-based estimator for $\psi$. Consequently, Theorem~\ref{thm:inference} justifies the use and asymptotic sharpness of the one-sided level-$\alpha$ confidence intervals yielded by Algorithm~\ref{alg1}, $$CI_N=[\widehat{\tau}-z_{1-\alpha}\widehat{\sigma},+\infty),$$ as well as the corresponding hypothesis tests. This ensures a valid, efficient and transparent inferential procedure for assessing combination therapies after adjusting for patient-level heterogeneity.

\begin{thm}[Asymptotic normality of $\psi$]
\label{thm:normality}
    Suppose Assumptions~\ref{assump:bliss}, \ref{assump:noise}, \ref{assump:unif-ext}, and~\ref{assump:match-bias} hold,  normalized matching estimator $\widehat \psi=\frac{1}{N_A} \sum_{T_{i} = A} \max\{ Y_{i}, Y_{\widetilde{i}} \}$ converges to a standard normal distribution,
    \begin{align*}
    \sqrt{N_A}\frac{(\widehat \psi - \psi)}{\widehat{\sigma}_{m} }\to_d N(0,1),
    \end{align*} where $\widehat{\sigma}_{m}$ is the sample deviation for $\{\max\{ Y_{i}, Y_{\widetilde{i}} \}:T_{i}=A\}$.
\end{thm}

\begin{thm}[Sharp inference for $\tau$]
\label{thm:inference}
    Suppose Assumptions~\ref{assump:bliss}, \ref{assump:noise}, \ref{assump:unif-ext}, and~\ref{assump:match-bias} hold, then the matching estimator $\widehat{\tau}$ is an asymptotically valid estimator for $\tau$, i.e. for a pre-specified level $\alpha$, given $CI_N$ and $p_N$ in Algorithm \ref{alg1}, \begin{align*}
        \liminf_{N\to\infty}P\left(\tau\in CI_N\right)&=1-\alpha\\
        \limsup_{N\to\infty}P(p_N\leq\alpha|H_0)&=\alpha.
    \end{align*}
\end{thm}

The core technical step in the proof of Theorems~\ref{thm:normality} and \ref{thm:inference} is to show that the matching estimator has a negligible error of order $o_P(N_A^{-1/2})$ relative to the oracle quantity $\max\{Y_{i}, Y_{i}^*\}$, where $Y_{i}^* = f_B(\mathbf{X}_{i}) + \epsilon_{\widetilde{i}}(B)$. Rather than relying on linearization or parametric modeling as in \citet{abadie2016matching,yang2023multiply}, we decompose the approximation error into two terms using an auxiliary inverse-weighting factor $\tilde{\pi}$. The first term is bounded using the bounded density of $\epsilon(A)-\epsilon(B)$ in Assumption~\ref{assump:noise} and the bias conditions in Assumptions~\ref{assump:unif-ext} and~\ref{assump:match-bias}, while the second term vanishes exactly by the construction of $\tilde{\pi}$. See the proof of Lemma~6 in the Appendix for more details.

\section{Robustness}\label{section:robustness}

In this section, we study the robustness of the proposed method when the identification condition in Section \ref{section:identification} is violated. Specifically, we relax Assumption~\ref{assump:bliss} and establish partial identification bounds for $\tau$ under non-negative dependence in Section~\ref{section:partial}, and provide a sensitivity analysis framework to quantify the impact of potential cross-world correlations in Section~\ref{section:sensitivity}. 

\subsection{Partial Identification}\label{section:partial}
In Section \ref{section:identification}, we establish identification of the proposed quantity of interest $\tau$, which measures the treatment effect of the combination therapy after adjusting for patient-level heterogeneity, under the conditional Bliss independence condition in Assumption~\ref{assump:bliss}. While widely considered in pharmacology, Assumption~\ref{assump:bliss} might be violated in practice; for instance, the combination of Dabrafenib and Trametinib can exhibit cross-resistance due to shared biological mechanisms \citep{palmer2017combination}. Here, we show that the quantity of interest $\tau$ remains partially identifiable even when Assumption ~\ref{assump:bliss} is violated, if a mild condition about positive cross-world correlation is satisfied as stated in Assumption~\ref{assump:nnc}.

\begin{asm}[Non-negative dependence measure]\label{assump:nnc}
$Cov[\mathbb{I}_{Y(A)> t},\mathbb{I}_{Y(B)> t}|\mathbf{X}]\geq 0$, for all $t\in\mathbb{R}$ a.s..
\end{asm}

Assumption~\ref{assump:nnc} requires that the dependence measure conditional on $\mathbf{X}$ is non-negative. This assumption is naturally satisfied under cross-resistance in pharmacology, such as nivolumab and ipilimumab \citep{palmer2017combination}. When the dependence measure in Assumption~\ref{assump:nnc} equals zero, it degenerates exactly to the conditional Bliss independence (Assumption~\ref{assump:bliss}), thereby unifying the two scenarios, i.e. zero dependence (conditional Bliss independence) and cross-resistance, under a unified framework.  Under Assumption~\ref{assump:nnc}, the partial identification and the validity of the inference on $\tau$ are given below in Theorems~\ref{thm:partial-id-H} and~\ref{thm:valid inference} respectively.

\begin{thm}[Partial identification]
\label{thm:partial-id-H}
Under Assumption~\ref{assump:nnc}, the following inequality holds: $\psi_H \geq \psi$, and $\tau_H \leq \tau$, and thus $\tau$ is partially identifiable. Specifically, the identity holds when Assumption~\ref{assump:bliss} holds.
\end{thm}
\begin{thm}[Valid inference for $\tau$]\label{thm:valid inference}
Suppose Assumptions~\ref{assump:noise}, \ref{assump:unif-ext}, \ref{assump:match-bias} and~\ref{assump:nnc} hold, then the matching estimator $\widehat{\tau}$ is an asymptotically valid estimator for $\tau$, i.e.\ for a pre-specified level $\alpha$, given $CI_N$ and $p_N$ in Algorithm~\ref{alg1},
\begin{align*}
    \liminf_{N\to\infty}P\left(\tau\in CI_N\right) &\geq 1-\alpha\\
    \limsup_{N\to\infty}P(p_N\leq\alpha\mid H_0) &\leq \alpha.
\end{align*}
\end{thm}

Theorem~\ref{thm:partial-id-H} establishes inequalities linking the non-point-identified parameters $\psi$ and $\tau$ to their partially identified bounds, which are relatively tight as discussed in Appendix~\ref{section:partial}. Theorem~\ref{thm:valid inference} then sidesteps the lack of point identification, as our proposed matching algorithm guarantees valid inference for the partially identified $\tau_H$ and enables valid testing of the proposed framework in Eq. \eqref{basic2}.

\subsection{Sensitivity Analysis}\label{section:sensitivity}
In Section~\ref{section:partial}, we show that the proposed estimate provides valid but possibly conservative inference for the true combination effect $\tau$ under the non-negative cross-world dependence condition in Assumption~\ref{assump:nnc}. To investigate the possible conservatism of partial identification, we introduce a sensitivity analysis framework parameterized by the cross-world correlation under a Gaussian working model, though extensions to other working models are possible.

Specifically, we consider a Gaussian working model for the joint distribution of the potential outcomes $(Y(A), Y(B))$ conditional on $\mathbf{X}$ as stated in Assumption~\ref{assump:Gaussianity}, which is compatible with the noise structure introduced in Assumption~\ref{assump:noise}. Such a Gaussian working model is widely adopted for sensitivity analysis of continuous potential outcomes \citep{deresa2021semiparametric}, under which the true combination effect can be represented as a function $\tau(\rho)$ w.r.t. the constant cross-world correlation $\rho = Corr[Y(A),Y(B)\mid\mathbf{X}] = Corr[\epsilon(A), \epsilon(B)]$. Under the Gaussian working model in Assumption~\ref{assump:Gaussianity}, $\rho\ge 0$ implies the partial identification condition in Assumption~\ref{assump:nnc} (See Proposition 2 in the Appendix), and the identified bound $\tau_H = \tau(0)$. Therefore, we can characterize the sensitivity via the gap between the identified bound and the true effect, $\Delta(\rho) := \tau(\rho) - \tau(0)$, which is non-negative under Assumption~\ref{assump:nnc}. The sensitivity parameter is then defined as the smallest correlation $\rho^{*}$ at which a negative $\tau(0)$ no longer guarantees $\tau(\rho)\le 0$:
\begin{equation}\label{senpara}
\rho^{*} =\inf_{\rho\in[0,1]}\left\{ \Delta(\rho)+\tau(0) > 0 \right\}.
\end{equation}

The interpretation and calculation of the sensitivity parameter $\rho^*$ in Eq. \eqref{senpara} is as follows. When $\rho<\rho^{*}$, a negative $\tau(0)$ implies $\tau(\rho)\le 0$, indicating that the lack of sufficient evidence for a true pharmacological interaction remains robust. When $\rho\ge\rho^{*}$, it is possible to have $\tau(\rho)> 0$ alongside $\tau(0)\le 0$; thus, the initial interpretation becomes sensitive, suggesting that a genuine pharmacological interaction may potentially exist for this endpoint. Under Assumption~\ref{assump:Gaussianity}, we can obtain a closed form for $\Delta(\rho)$ based on the conditional means and variances as stated in Theorem~\ref{thm:Gaussianity}. Therefore, the sensitivity parameter $\rho^*$ can be directly estimated from the marginal distributions of the data.

\begin{asm}[Gaussianity]\label{assump:Gaussianity}
Conditionally on covariates $\mathbf{X} = \mathbf{x}$, the potential outcomes $(Y(A), Y(B))$ follow a bivariate normal distribution:
\begin{align*}
    (Y(A),\, Y(B)) \mid \mathbf{X} = \mathbf{x} \;\sim\;
\mathcal{N}\!\left(
\begin{pmatrix}
f_A(\mathbf{x}) \\[2pt]
f_B(\mathbf{x})
\end{pmatrix},
\begin{pmatrix}
\sigma_A^{2} & \rho\,\sigma_A\sigma_B \\[2pt]
\rho\,\sigma_A\sigma_B & \sigma_B^{2}
\end{pmatrix}
\right),
\end{align*}
where $\sigma_A$, $\sigma_B$, and $\rho$ are constants that do not depend on $\mathbf{X}$.
\end{asm}

\begin{thm}[Closed form of maximal treatment effect under Gaussianity]\label{thm:Gaussianity}
Suppose Assumptions~\ref{assump:noise} and~\ref{assump:Gaussianity} hold, and suppose $\rho\in[0,1]$. 
For each $\mathbf{x}\in\mathbb{X}$, the maximal treatment effect given $\mathbf{X}=\mathbf{x}$ is:
\[
\begin{aligned}
m(\rho,\mathbf{x})
&:=E[\max\{Y(A),Y(B)\}\mid\mathbf X=\mathbf x]\\
&= f_A(\mathbf{x})\,
   \Phi\!\!\left(
      \frac{f_A(\mathbf{x})-f_B(\mathbf{x})}
           {\sqrt{\sigma_A^{2}+\sigma_B^{2}-2\rho\sigma_A\sigma_B}}
   \right)
 + f_B(\mathbf{x})\,
   \Phi\!\!\left(
      \frac{f_B(\mathbf{x})-f_A(\mathbf{x})}
           {\sqrt{\sigma_A^{2}+\sigma_B^{2}-2\rho\sigma_A\sigma_B}}
   \right)\\
&\quad + \sqrt{\sigma_A^{2}+\sigma_B^{2}-2\rho\sigma_A\sigma_B}\,
   \phi\!\!\left(
      \frac{f_A(\mathbf{x})-f_B(\mathbf{x})}
           {\sqrt{\sigma_A^{2}+\sigma_B^{2}-2\rho\sigma_A\sigma_B}}
   \right),
\end{aligned}
\]
where $\Phi$ and $\phi$ denote the standard normal CDF and PDF, respectively.
The distance between two maximal treatment effects is:
\[
\begin{aligned}
\Delta(\rho)
&:=\psi(0)-\psi(\rho)
= \int_{\mathbb{X}}
   \bigl(m(0,\mathbf{x})-m(\rho,\mathbf{x})\bigr)\,
  \mathrm dP_{\mathbf{X}}(\mathbf{x}),
\end{aligned}
\]
where $P_{\mathbf{X}}(\mathbf{x})$ is the probability measure induced by covariates $\mathbf{X}$.
\end{thm}

The Gaussian sensitivity analysis may be extended to non-Gaussian settings using standard copula models \citep{bodik2025cross}. Specifically, one may specify a parametric copula $C_\theta$ to model the joint cross-world dependence between $Y(A)$ and $Y(B)$ given $\mathbf{X}$ while preserving the data-evidenced marginals, where the copula parameter $\theta$ plays the role of the sensitivity parameter analogous to $\rho$ in the Gaussian case. Under a concordance-ordered copula family containing an independence member $\theta_0$, $\Delta(\theta) := \tau(\theta_0) - \tau(\theta)$ remains well-defined and monotonic under this broader framework. For instance, one may evaluate diverse joint tail behaviors and dependence structures using parametric copulas such as the Frank copula. We focus on the Gaussian working model in the present paper, as it yields a directly interpretable sensitivity parameter and admits transparent closed-form results; exploration of richer copula specifications is left for future work.

\section{Simulation}\label{section:sim}
Consider the simulation settings where the total sample size is $N = N_A + N_B + N_{A+B} \in \{300, 500, 1000\}$. The treatment assignment probabilities are specified as $P(T = A) = 0.25$, $P(T = B) = 0.5$, and $P(T = A+B) = 0.25$. The covariate $\mathbf{X}=(X_1,X_2)^\top$ has a two-dimensional independent $\mathrm{Uniform}(0,1)$ components, and the potential outcomes follow the model $Y(T)=f_T(\mathbf{X})+\epsilon(T)$, where $f_A(\mathbf{X})=100(X_1+X_2)$, $f_B(\mathbf{X})=30(X_1+X_2)$ and $f_{A+B}(\mathbf{X})=10.5+90(X_1+X_2)$. The errors $(\epsilon(A),\epsilon(B))$ are bivariate Gaussian with mean zero, standard deviation $\sigma=25$ which is of similar scale to that in the real data analysis, and correlation $\rho$, while $\epsilon(A+B)\sim\mathcal{N}(0,\sigma^2)$ is independent. We vary the correlation parameter $\rho \in \{0, 0.15, 0.30, 0.45, 0.60\}$ to study the validity and robustness of our method under different dependence structure between $Y(A)$ and $Y(B)$ as indicated in Table~\ref{tab:coverage-MB-merged}. 


We compare the proposed method with the classical evaluation framework. For the proposed procedure, we use a second-order power series as the outcome estimator for $f_B(\mathbf{X})$, trained on a separate, independent sample with an identical sample size $N$, and the optimal pair-matching algorithm is implemented in the \texttt{optmatch} package in R, with matching performed without replacement. Confidence intervals are constructed using the proposed matching-based standard errors in Algorithm \ref{alg1}. For the classical test in \eqref{basic}, we use the debiased-bootstrap procedure proposed by \citet{guo2021inference}. Specifically, resampling with $N$ bootstrap replications is performed separately within the three treatment groups ($A$, $B$, and $A+B$), and the bias induced by the $\max$ operator in Eq. \eqref{basic} is addressed through the debiasing step. The resulting one-sided confidence interval serves as the classical benchmark. Table~\ref{tab:coverage-MB-merged} reports the empirical coverage probabilities of the nominal 95\% confidence intervals by $10^5$ repetitions under Settings~1--5 with different $\rho$. Specifically, in Settings 1--3, $\eta=E[Y(A+B)]-\max[E[Y(A)],E[Y(B)]]$, and $\tau_H \le \tau < 0 < \eta$, meaning that the combination's treatment effect exceeds the better monotherapy in marginal mean but does not exceed the patient-level best monotherapy benchmark as discussed in Eq. \eqref{inequal}.

\begin{table}[H]
\centering
\caption{Empirical coverage probabilities of nominal 95\% confidence intervals for $\tau$ and $\tau_H$ based on the proposed and classical procedures with standard error below 0.001.}
\label{tab:coverage-MB-merged}
\scriptsize
\setlength{\tabcolsep}{3.2pt}
\renewcommand{\arraystretch}{1.08}
\begin{tabular}{@{}l*{4}{cccc}@{}}
\toprule
& \multicolumn{3}{c}{} & \multicolumn{4}{c}{\textbf{$N=300$}} & \multicolumn{4}{c}{\textbf{$N=500$}} & \multicolumn{4}{c}{\textbf{$N=1000$}} \\
\cmidrule(lr){5-8}\cmidrule(lr){9-12}\cmidrule(lr){13-16}
\textbf{Setting} & \multicolumn{3}{c}{\textbf{True Values}} & \multicolumn{2}{c}{\textbf{Proposed}} & \multicolumn{2}{c}{\textbf{Classical}} & \multicolumn{2}{c}{\textbf{Proposed}} & \multicolumn{2}{c}{\textbf{Classical}} & \multicolumn{2}{c}{\textbf{Proposed}} & \multicolumn{2}{c}{\textbf{Classical}} \\
\cmidrule(lr){2-4}\cmidrule(lr){5-6}\cmidrule(lr){7-8}\cmidrule(lr){9-10}\cmidrule(lr){11-12}\cmidrule(lr){13-14}\cmidrule(lr){15-16}
& $\tau_H$ & $\tau$ & $\eta$ & $\tau$ & $\tau_H$ & $\tau$ & $\tau_H$ & $\tau$ & $\tau_H$ & $\tau$ & $\tau_H$ & $\tau$ & $\tau_H$ & $\tau$ & $\tau_H$ \\
\midrule
1 $(\rho=0)$    & -0.690 & --      & 0.471 & 0.941 & 0.941 & 0.855 & 0.855 & 0.950 & 0.950 & 0.853 & 0.853 & 0.953 & 0.953 & 0.835 & 0.835 \\
2 $(\rho=0.15)$ & -0.690 & -0.448 & 0.471 & 0.947 & 0.940 & 0.862 & 0.858 & 0.952 & 0.948 & 0.866 & 0.859 & 0.957 & 0.948 & 0.851 & 0.834 \\
3 $(\rho=0.30)$ & -0.690 & -0.210 & 0.471 & 0.949 & 0.943 & 0.872 & 0.857 & 0.958 & 0.947 & 0.878 & 0.858 & 0.960 & 0.948 & 0.871 & 0.833 \\
4 $(\rho=0.45)$ & -0.690 & 0.006 & 0.471 & 0.953 & 0.948 & 0.876 & 0.854 & 0.964 & 0.945 & 0.884 & 0.857 & 0.965 & 0.949 & 0.881 & 0.836 \\
5 $(\rho=0.60)$ & -0.690 & 0.196 & 0.471 & 0.956 & 0.949 & 0.884 & 0.859 & 0.965 & 0.943 & 0.886 & 0.858 & 0.968 & 0.952 & 0.886 & 0.838 \\
\bottomrule
\end{tabular}
\medskip
\begin{minipage}{0.98\textwidth}
\footnotesize
\textit{Notes:} Here, $\eta=E[Y(A+B)]-\max[E[Y(A)],E[Y(B)]]$. "Proposed" refers to the proposed matching-based procedure in Algorithm \ref{alg1}, whereas "Classical" refers to the debiased-bootstrap procedure of \citet{guo2021inference}. The symbol "--" indicates that $\tau$ is not separately reported when $\rho=0$.
\end{minipage}
\end{table}


From Table \ref{tab:coverage-MB-merged}, we see that the empirical coverage probability of the classical method for $\tau$ is clearly below the nominal level 95\%. This is because $E[Y(A{+}B)] - E[\max\{Y(A), Y(B)\}] < E[Y(A{+}B)] - \max\{E[Y(A)], E[Y(B)]\}$, leading the classical interval fail to cover the true target parameter $\tau=E[Y(A{+}B)] - E[\max\{Y(A), Y(B)\}]$. The lack of coverage demonstrates the inadequacy of classical evaluation, which may attribute marginal superiority to pharmacological interaction, even when the combination does not exceed the patient-level best monotherapy benchmark (Settings 1--3). It also explains why the classical evaluation can be overly optimistic in assessing the additional value of the combination, because comparable outcomes may in principle be achieved under the oracle patient-level monotherapy benchmark. On the contrary, our matching-based estimator achieves the nominal level well, though it is slightly conservative when $\rho>0$ and $\tau$ is partially identifiable (Settings~2--5). This demonstrates that the proposed inference procedure can evaluate the combination therapy appropriately and robustly after adjusting for patient-level heterogeneity.  

To further demonstrate the generalizability and robustness of our approach, we include additional simulation scenarios in the Appendix (Settings~A.6--A.10). These supplementary settings vary the outcome estimator (including linear regression, neural networks, and Lasso), the data source for outcome estimation (independent external data versus internal data reuse), the matching algorithm (nearest-neighbor versus optimal matching), noise scale ($\sigma=25$ versus $\sigma=1$) and the covariate dimension (extending up to $p = 35$). Across these diverse scenarios, the results are robust across parametric and nonparametric models, internal and external estimation, optimal and other standard matching, lower- and higher-noise as well as lower- and higher-dimensional settings. These consistent findings empirically support the validity and robustness of the proposed method for inference on the partially identified bounds across a broad range of data-generating processes.

\section{Application: Re-analysis of ACTG 175 Trial}\label{section:application}

In this section, we apply our proposed method to the AIDS Clinical Trials Group (ACTG) 175 trial. We investigate the combination therapy that consists of $A=\text{zidovudine}$ and $B=\text{didanosine}$, and thus $A+B=\text{zidovudine+didanosine}$, and the outcome is selected as CD4 cell count at 20 weeks (cd420). For covariate $\mathbf{X}$, we consider age in years at baseline (age), weight in kg at baseline (wtkg), Karnofsky score (karnof), and number of days of previously received antiretroviral therapy (preanti). In our analysis, the sample sizes for the zidovudine arm, the didanosine arm, and the combination therapy group are 532, 561, and 522, respectively. 

We compare the classical evaluation in Eq.~\eqref{basic} with our proposed evaluation in Eq.~\eqref{basic2}. For the classical evaluation, we apply the selected subgroup test of \citet{guo2021inference}, implemented via a debiased bootstrap. For the proposed evaluation, we use optimal matching without replacement. The working outcome model is estimated using a second-order power-series regression; see \citet{abadie2011bias,lin2023estimation}. In particular, $\widehat f_B(\mathbf X)$ is estimated directly using observations from the didanosine arm, without an external dataset.

\begin{table}[H]
\centering
\caption{Evaluation Results of Zidovudine plus Didanosine based on the ACTG 175 trial under the classical and proposed frameworks.}
\label{tab:testing-results-merged}
\setlength{\tabcolsep}{14pt} 
\renewcommand{\arraystretch}{1.1}
\begin{tabular}{@{}l ccc@{}}
\toprule
\textbf{Procedure} & \textbf{Matched Pairs} & \textbf{Estimate} & \textbf{One-Sided CI Lower Bound} \\
\midrule
Classical & --    & $28.848$  & $17.785$ \\
Proposed  & $532$ & $-30.606$ & $-45.357$ \\
\bottomrule
\end{tabular}
\medskip
\begin{minipage}{0.98\textwidth}
\small 
\textit{Notes:} "Classical" refers to the test of the hypothesis in Eq.~\eqref{basic}, where inference is conducted via the debiased-bootstrap procedure \citep{guo2021inference}. "Proposed" refers to the matching-based test of the hypothesis in Eq. \eqref{basic2} via Algorithm \ref{alg1}. The symbol "--" indicates not applicable.
\end{minipage}
\end{table}

The two methods yield two different results as summarized in Table \ref{tab:testing-results-merged}. Under the classical marginal comparison, the positive estimate and the one-sided lower confidence bound exceeding zero suggest that the combination therapy outperforms the better monotherapy in terms of the marginal mean outcome. However, we must note that indeed such a classical evaluation alone does not distinguish drug interaction from a benefit that can be explained by patient-level heterogeneity as we discuss before.

In contrast, under our proposed matching evaluation, the negative estimate and the one-sided lower confidence bound well below zero provide no evidence that the combination therapy outperforms the best individual-level monotherapy response once patient-level heterogeneity is accounted for. Taken together, these results suggest that, for the 20-week CD4 outcome considered here, the apparent superiority observed in the traditional marginal comparison is compatible with patient-level independent action and provides no evidence of additional interaction benefit beyond the best individual-level monotherapy response after adjusting for patient-level heterogeneity.

\begin{figure}[H]
    \centering
    \includegraphics[width=0.75\linewidth]{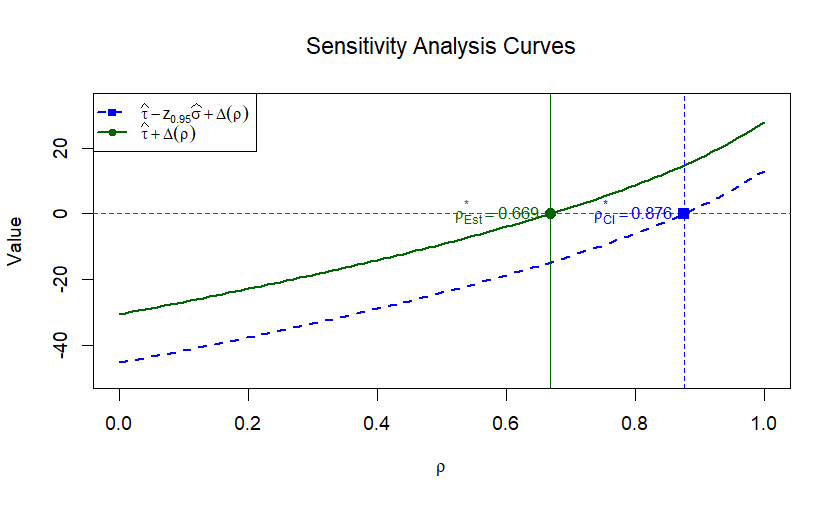}
    \caption{Sensitivity analysis for the ACTG 175 trial: the sensitivity parameters for the point estimate ($\rho^*_{Est}$) and confidence interval  ($\rho^*_{CI}$) are 0.669 and 0.876 respectively.}
    \label{fig:sensitivity_ACTG}
\end{figure}

To study the robustness of the analysis, we conduct a sensitivity analysis following the Gaussian framework described in Section~\ref{section:sensitivity} for both the point estimate and the confidence interval. From the resulting curves (Figure~\ref{fig:sensitivity_ACTG}), the sensitivity parameters for the point estimate and the confidence interval are $0.669$ and $0.876$, respectively, both of which represent relatively large cross-world correlation thresholds. This suggests that the observed benefit of zidovudine plus didanosine remains largely compatible with patient-level heterogeneity, providing no sufficient evidence of a genuine interaction effect unless an exceptionally large cross-world correlation exists for CD4 outcome. A model check is also provided in the Appendix, which supports the plausibility of the bivariate normal sensitivity analysis framework in the ACTG 175 trial.

\section{Discussion}\label{section:discussion}

This work provides a robust and model-free evaluation framework for the expected outcome of combination therapy under patient-level heterogeneity, thereby distinguishing the benefits of independent drug action from true pharmacological interactions. By establishing (partial) identifiability and addressing the inherent non-linearity via an outcome-based optimal-matching scheme, our approach ensures valid and transparent statistical inference with a $\sqrt{N}$-rate of convergence for the cross-world target parameter, thereby preventing the overly optimistic or inadequate conclusions often drawn by classical evaluation that neglects patient-level heterogeneity.

Several interesting directions might warrant future research. First, beyond the two-agent case studied here, extension of the current framework to higher-order combinations involving three or more agents bears both methodological and practical importance. Second, apart from the continuous outcome studied here, combination therapies might be evaluated using censored time-to-event endpoints (e.g., PFS and OS) or count outcomes in practice, and adapting the current maximal-response functional and its asymptotic theory to these diverse endpoints is of great importance. In the end, in addition to the max-type causal quantity considered here, the proposed matching methods and the associated theories might be extended to other non-linear causal functionals of potential outcomes, where further investigations are needed.

\bigskip

\section*{Acknowledgements}
We would like to thank Florence Bourgeois, Tianxi Cai, Adam Palmer, Deborah Plana, Hanyu Wei, Jiayue Xu, Yuming Zhang and Yunhao Zheng for helpful suggestions and discussion about the project. The work was partially supported by grants from Research Grants Council of the Hong Kong Special Administrative Region, China (HKUST~26308323, HKUST~16310125 and Hong Kong PhD Fellowship Scheme), the Seed fund of the Big Data for Bio-Intelligence Laboratory (Z0428) and the grant L0438 from the Hong Kong University of Science and Technology. 

\section*{Data Availability Statement}

The ACTG~175 data that support the findings in Section~\ref{section:application} of this paper are publicly available as part of the R package \texttt{speff2trial} on the Comprehensive R Archive Network (\url{https://cran.r-project.org/package=speff2trial}). Simulation code reproducing the numerical results in Section~\ref{section:sim} is provided as Supplementary Material.

\section*{Supplementary Material}
Supplementary Material for Appendix and the R code to reproduce the numerical results referenced in Sections~\ref{section:sim}, and \ref{section:application} are available online.


\bibliographystyle{apalike}
\bibliography{refs}

\end{document}